\documentclass[aip,apl,amsmath,amssymb,reprint]{revtex4-1}

\usepackage{graphicx}
\usepackage{dcolumn}
\usepackage{bm}

\usepackage[utf8]{inputenc}
\usepackage[T1]{fontenc}
\usepackage{mathptmx}
\usepackage{etoolbox}
\usepackage{xcolor}

\usepackage{graphicx}
\usepackage[caption=false]{subfig}

\newcommand{\mrm}[1]{\mathrm{#1}}

\makeatletter
\def\@email#1#2{%
 \endgroup
 \patchcmd{\titleblock@produce}
  {\frontmatter@RRAPformat}
  {\frontmatter@RRAPformat{\produce@RRAP{*#1\href{mailto:#2}{#2}}}\frontmatter@RRAPformat}
  {}{}
}%
\makeatother
\begin{document}

\preprint{AIP/123-QED}

\title[]{Sound diffusion with spatiotemporally modulated acoustic metasurfaces}
\author{Janghoon Kang}
 \email[]{jh3010.kang@utexas.edu}
\affiliation{Walker Department of Mechanical Engineering, The University of Texas at Austin, Austin, TX, USA, 78712-1591}
\author{Michael R. Haberman}
 \email[]{haberman@utexas.edu}
\affiliation{Walker Department of Mechanical Engineering, The University of Texas at Austin, Austin, TX, USA, 78712-1591}

\date{\today}

\begin{abstract}
Traditional sound diffusers are quasi-random phase gratings attached to reflecting surfaces whose purpose is to augment the spatiotemporal incoherence of the acoustic field scattered from reflective surfaces. This configuration allows one to cover a large reflecting surface by periodically tiling the diffuser unit cells to cover a large area while reducing undesirable specular reflection for incident plane waves. However, the periodic arrangement of the unit cells leads to coherent constructive and destructive interference in the scattered field in some directions which is undesirable for achieving acoustic diffusivity. The spatial uniformity of acoustic energy scattered from conventional diffusers constructed in this way is a fundamental limitation of the traditional approach which is not easily overcome when one wishes to cover large reflecting surfaces. In this work, we investigate spatiotemporal modulation of the surface acoustic admittance of an acoustic diffuser as a new approach to improve the sound diffusion. We develop a semi-analytical model that employs Fourier series expansion to determine the scattered sound field from a surface admittance consisting of a quadratic residue diffuser (QRD) design whose individual well admittances are modulated with a traveling wave with modulation frequency, $\omega_{\mrm{m}}$, amplitude, $Y_\mrm{m}$, and a wavenumber that matches the unit cell length, $\Lambda$. We observe significant improvement in diffusion performance due to the fact that the spatiotemporal modulation scatters sound into additional frequency--wavenumber pairs associated with harmonics of the modulation frequency and their diffraction orders. The semi-analytical model results are verified using time-domain finite element model.
\end{abstract}
\maketitle



Conventional sound diffuser designs employ surface irregularity, usually in the form of acoustic cavities\cite{Schroeder1975, Schroeder1979, d1984reflection, Cox1994}, to maximize the randomness in the phase of the acoustic field scattered from each point on the diffuser. Randomization in the scattered field results in a minimization in the angular dependence of the scattered field intensity when subjected to plane wave incidence, thereby increasing the diffusivity\cite{ballestero2019experimental}. Traditional diffuser unit cell designs are then tiled over larger areas using a periodically repeating arrangement in space. As a result, grating lobes are generated in the field scattered from the collection of units, which significantly degrades acoustic diffusion performance\cite{angus2000using, cox2003schroeder, cox2016acoustic}. Conventional diffuser performance can therefore be improved by modifying the design to disrupt periodicity in unit cell tiling, for example, by adding different types of sequences\cite{angus2000using,cox2016acoustic}. However, these approaches require increased diffuser size to introduce additional elements or additional depth to create a quasi-random geometric sequences with sufficiently different phase distribution. This work presents an unique approach to enhance acoustic diffusion using spatiotemporal modulation of the input impedance of traditional diffuser design geometries. We show that this approach greatly improves diffuser performance while maintaining the geometry of traditional diffuser designs such as the quadratic residue diffuser (QRD), first introduced by Schroeder\cite{Schroeder1979}.

Spatiotemporal modulation of material properties has recently gained interest as a mechanism to generate nonreciprocal wave propagating in the electromagnetic, acoustic, and elastic wave propagation thereby enabling new avenues to control wave fields \cite{nassar2020nonreciprocity}. Early work considered the effects of spatiotemporal modulation in the form of a pump wave imposed by an external source in an unbounded medium on electromagnetic wave propagation\cite{cassedy1963dispersion}, while more recent research studied spatiotemporal modulation of stiffness or density in the form of a traveling wave to induce nonreciprocal propagation in acoustic and elastic domains\cite{trainiti2016non, nassar2017non, nassar2017modulated, vila2017bloch, riva2019generalized, trainiti2019time, goldsberry2020nonreciprocal, sugino2020nonreciprocal, shen2019nonreciprocal, li2019nonreciprocal, zhu2020non}. Modulation has also been employed as a means to mimic symmetry-breaking momentum bias present in fluid-flow\cite{fleury2014sound} in order to create nonreciprocal acoustic and electromagnetic circulators\cite{wang2005magneto, fleury2015subwavelength, goldsberry2022nonreciprocity}. Modulation has also been used in metasurfaces to induce nonreciprocal reflection and transmission of elastic and acoustic waves\cite{zhu2020non, adlakha2020frequency} as well phase shifting \cite{wang2020theory}, total power combining\cite{wang2021space}, nonreciprocal scattering\cite{guo2019nonreciprocal}, and the Doppler-like frequency shift of reflected optical waves \cite{shaltout2015time, shaltout2019spatiotemporal}.

Spatiotemporal modulation of bounded and unbounded media leads to the generation of harmonics in propagating and forced standing wave fields, respectively, at multiples of the modulation frequency and wavenumber\cite{vila2017bloch, wallen2019nonreciprocal, nassar2017non, goldsberry2019non}. While the presence of these harmonics can complicate the design of nonreciprocal devices\cite{goldsberry2020nonreciprocal}, the generation of additional frequencies and wavenumbers is highly useful if one wishes to efficiently diffuse acoustic energy in time and space. The idea of scattering energy into multiple frequencies and wavelengths via nonlinear interactions is a well-known approach to efficiently absorb acoustic\cite{konarski2020acoustic} and vibrational energy\cite{motato2017targeted}, but this concept has not been considered as a means to improve acoustic diffusion. Further, the use of spatiotemporal modulation to scatter energy into different frequencies does not require nonlinear behavior or high amplitude signals. In this case, the surface input impedance becomes a linear, time varying parameter for every point no the surface, which requires a generalization of the conventional grating equation to model the diffuser performance.

In this work, we adapt the approach that Goldsberry et al.\cite{goldsberry2020nonreciprocal} employed to investigate nonreciprocal vibrations in Euler beams to the present case of surface acoustic admittance modulation of a traditional Schroeder diffuser scheme to improve sound diffusion. We derive a theoretical model that considers a spatiotemporally modulated surface admittance in the form of a traveling wave with modulation frequency, $\omega_\mrm{m}$, amplitude, $Y_\mrm{m}$, and a wavenumber that matches the unit cell length, $\Lambda$. With the properly selected modulation parameters, the far-field scattering is determined using a semi-analytical model shows that the directions of the diffraction modes are rotated from the original ones due to the addition of wavenumber and frequency harmonics which do not have symmetric amplitudes with respect to the incident frequency and wavenumber. Further, we show that the diversity of scattered wavenumbers generated by the modulation leads to additional diffracted modes, thereby reducing the prominence in grating lobes of the classical diffuser design and significantly improving diffuser performance. These results demonstrate that the spatiotemporal modulation of the phase grating on a sound diffuser is a new approach to tailor the scattered sounds and to improve the diffusion performance in order to disperse scattered sound power.\\

\begin{figure}
    \includegraphics[width=0.49\textwidth]{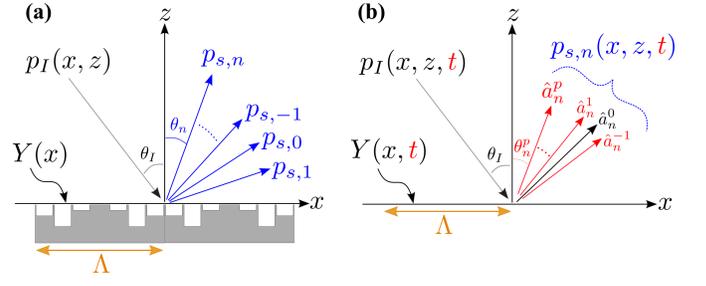}
    \caption{\label{fig:scattering} Diagram of the incident and scattered sound in front of a sound diffuser. The scattered field is composed of plane-waves reflected from (a) an unmodulated surface, or (b) a modulated surface.}
\end{figure}

Consider the classical, Schroeder-type one-dimensional (1D) sound diffuser shown in Fig.~\ref{fig:scattering}a, where the local input admittance $Y(x)$ is a function of the depth of the cavity at the position. The position-dependent admittance can be determined using various approaches to optimize sound diffusion\cite{Cox1994}. We have used a QRD with seven cavities for this demonstration, but the approach presented here can accommodate any admittance profile. We consider a generalized case of a diffuser shown in Fig.~\ref{fig:scattering}b where the local input admittance is decomposed into a static and modulated components as $Y(x) = Y_{0}(x) + Y_\mrm{m}(x,t)$, where $Y_{0}(x)$ is the surface admittance of the unmodulated sound diffuser and $Y_\mrm{m}(x,t)$ is the modulation. The scattering problem is solved by decomposing the sound field into incident and scattered fields which are represented by the summation of the diffraction modes\cite{berkhout1979theory,mechel1995wide,wu2000profiled} as shown in Eq.~\ref{decomp}. To account for the modulation in the analytical model, the acoustic field and the acoustic surface admittance of the diffuser are expressed as the function of time and space\cite{goldsberry2019non}. The total acoustic field is therefore described as
\begin{align}
    p\left(x,z,t\right)=&\, p_\mrm{I}\left(x,z,t\right)+p_\mrm{s}\left(x,z,t\right)\nonumber \\
    =&\, a_\mrm{I}e^{j\left(\omega_{0}t-k_{x}x+k_{z}z\right)}+\sum_{n=-\infty}^{\infty}a_{n}(t)e^{-j\left(\beta_{n}x+\gamma_{n}z\right)},
    \label{decomp}
\end{align}
where $p_\mrm{I}\left(x,z,t\right)$ and $p_\mrm{s}\left(x,z,t\right)$ represent the incident and the scattered sound respectively, $\omega_0$ is the frequency of the incident plane wave, $k_{x}=k\sin\theta_\mrm{I}$ and $k_{z}=k\cos\theta_\mrm{I}$ are the wave numbers of the incident sound in the $x$ and $z$-direction, $\theta_\mrm{I}$ is the angle of incidence, $\beta_{n}$ and $\gamma_{n}$ are the wave numbers of the $n^\mrm{th}$ diffraction mode in the $x$- and $z$-direction, respectively, and $\theta_{n}$ is the angle of the propagation of the $n^\mrm{th}$ diffracted mode. The scattering amplitudes, $a_{n}(t)$, are represented as the function of time to consider the effect of the temporal modulation. By using the classical grating equation, $\beta_{n}$ can be related to the geometry of the sound diffuser as
\begin{align}
    \beta_{n}=k_{x}+\frac{2\pi}{\Lambda}n,
    \label{beta}
\end{align}
where $\Lambda$ is the length of one period of the acoustic diffuser.

The momentum equation at the diffuser interface, Eq.~\ref{momentum_imp}, must consider the interaction of the incident field with the spatiotemporally modulated surface impedance through the normal component of the particle velocity, $u_{z}(x,z=0,t)$ and the acoustic pressure, $p(x,z=0,t)$. Using the definition of the surface impedance, the normal velocity can be related to the sound pressure at $z=0$ and the surface impedance,  $Z(x,t)$.
\begin{align}
    \rho_{0}\frac{\partial u_{z}}{\partial t}\Big|_{z=0}=-\frac{\partial p}{\partial z} \Big|_{z=0},\quad
    u_{z}(x,t)|_{z=0} = - \frac{p(x,t)}{Z(x,t)}\Big|_{z=0}.\label{momentum_imp}
\end{align}
The sound pressure, normal velocity, and surface impedance must all be expressed as the function of time and space to account for spatiotemporal modulation. The normalize surface admittance, $Y(x,t) = \rho_0 c_0 / Z(x,t)$, where $\rho_0 c_0$ is the characteristic impedance of the background medium, is represented by the expansion of the harmonic terms in the form of the admittance
\begin{align}
    Y(x,t)=Y_{0}(x)+Y_\mrm{m} \sum_{\substack{p=-\infty \\ p\neq0}}^{\infty} \hat{\chi}^{p}(x)e^{jp\omega_\mrm{m}t},\label{admit}
\end{align}
where $Y_\mrm{m}$ and $\omega_\mrm{m}$ are the modulation magnitude and the modulation frequency and $\hat{\chi}^{p}(x)$ represents the spatial variation of the modulation amplitude due to the $p^\mrm{th}$ harmonic component of the modulation function. In order to conserve momentum in the normal direction in the presence of spatiotemporal modulation, the surface admittance as expressed by Eq.~\ref{admit} requires the introduction of modulation harmonics in the scattered wave field. The complex amplitude of the scattered field must therefore be expressed with a Fourier expansion of the harmonics of the modulation frequency. Specifically, the the scattering amplitudes $a_{n}(t)$ in Eq.~(\ref{decomp}) are represented as
\begin{align}
    a_{n}(t) = \sum_{p=-\infty}^{\infty}\hat{a}_{n}^{p}e^{j\left(\omega_{0}+p\omega_\mrm{m}\right)t}=\sum_{p=-\infty}^{\infty}\hat{a}_{n}^{p}e^{j\omega_{p}t},
    \label{amp_harmonic}
\end{align}
where $\hat{a}_{n}^{p}$ represents the amplitudes of $p^\mrm{th}$ modulation harmonic and $\omega_{p} = \omega_{0} + p\omega_\mrm{m}$ which is the scattered frequency shifted by the integer harmonic of the modulation frequency.
The wavenumber of the scattered sound in the $z$-direction can also be expressed using the shifted frequency, $\omega_{p}$, as
\begin{equation}
    \gamma_{n}^{p} = \sqrt{\left(\frac{\omega_{p}}{c_{0}}\right)^2 - \beta_{n}^2}.
    \label{gamma}
\end{equation}
Furthermore, the field of each diffraction order is scattered at the angle
\begin{equation}
    \theta_{n}^{p} = \sin^{-1}{\left(\frac{\beta_{n}}{\omega_{p}/c_{0}}\right)}= \sin^{-1}{\left(\frac{\sin{\theta_\mrm{I}}+\frac{\lambda_{0}}{\Lambda}n}{1+p\frac{\omega_{m}}{\omega_{0}}}\right)},
    \label{direction}
\end{equation}
where $n$ is the diffraction order associated with the diffuser periodicity and $p$ is the modulation harmonic index. When spatiotemporal modulation is absent, $p=0$ and $\omega_\mrm{m}=0$, and Eq.~(\ref{gamma}) and Eq.~(\ref{direction}) simplify to the standard grating formula. The shifted frequency of the scattered sound, $\omega_{p}$, caused by the modulation clearly affects the directions of each diffraction mode by shifting them from what one would observe with with an unmodulated diffuser surface. We clearly observe that the primary effect of the spatiotemporal modulation on the scattered sound is to scatter energy into modulation harmonics and their associated diffraction angles which leads to significant improvement of the diffusion performance as shown below.
Finally, we note that diffraction mode having the purely imaginary $\gamma_{n}^{p}$ are evanescent and cannot propagate into the far-field. Propagating diffraction modes are therefore bounded by the relations
\begin{equation}
    -\frac{\Lambda}{\lambda_{0}}\left(1+\sin{\theta_\mrm{I}}+p\nu\right) \leq n \leq \frac{\Lambda}{\lambda_{0}}\left(1-\sin{\theta_\mrm{I}}+p\nu\right),
    \label{prop_modes}
\end{equation}
where $\nu:=\frac{\omega_{m}}{\omega_{0}}$. Equations (\ref{decomp})--(\ref{prop_modes}) allow the calculation of the scattering amplitudes $\hat{a}_{n}^{p}$, for the entire range of the index $n$ and $p$ which represent the diffraction modes and the harmonics of the modulation frequency, respectively, given the incident amplitude and frequency\cite{supplement}. Once the $\hat{a}_{n}^{p}$ are obtained, the scattered sound pressure on the surface of the sound diffuser can be calculated using Eqs.~(\ref{decomp}) and (\ref{amp_harmonic}). We then can then employ angular spectrum methods approximation to estimate the far-field directivity using the non-evanescent components of the scattering amplitudes at $z=0$\cite{williams1999fourier}.\\

\begin{figure}
    \includegraphics[width=0.5\textwidth]{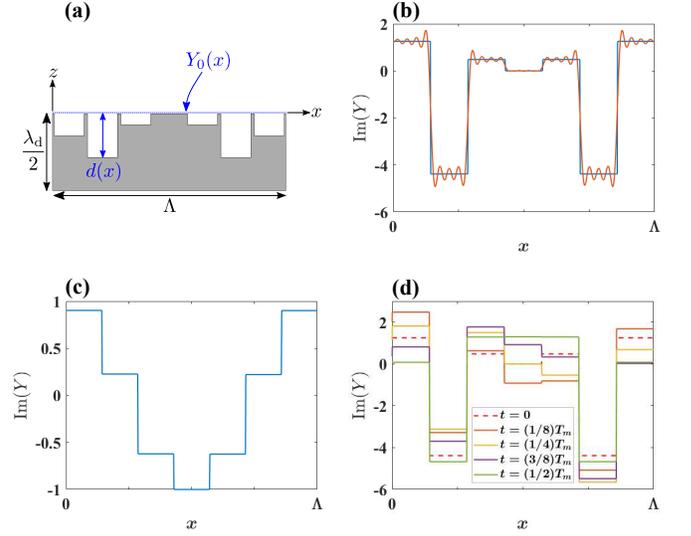}
    \caption{\label{fig:admittance}(a) Diagram of the original QRD with seven cavities in one period. (b) Imaginary part of $Y_{0}(x)$. Exact equation (solid blue line) and Fourier expansion with $N=14$ (solid orange line). (c) Spatial modulation function at $t = 0$. (d) Surface acoustic admittance as a function of space for different instants in time.}
\end{figure}

To demonstrate the ability spatiotemporal modulation to improve sound diffuser performance, consider two periods of a one dimensional QRD with seven cavities and design frequency of 5~kHz as a baseline model. The cavity depths (left to right) of the design shown in Fig.~\ref{fig:admittance}a are $4.9\mrm{mm}, 19.6\mrm{mm}, 9.8\mrm{mm}, 0\mrm{mm}, 9.8\mrm{mm}, 19.6\mrm{mm}, 4.9\mrm{mm}$. The surface admittance in Fig.~\ref{fig:admittance}a is approximated as $Y_{0}(x) = j\tan{\left(k_{0}\,d(x)\right)}$ because each cavity of the QRD can be considered as the closed-end tube. Figure \ref{fig:admittance}b shows the normalized surface admittance $Y_{0}(x)$ when assuming negligible wall thickness and that the surface admittance is assumed to be constant across the opening of each cavity. The modulation function is similarly discretized in space and modulated in time and space using Eq.\ref{admit_modfunction} in Fig.~\ref{fig:admittance}c. Figure~\ref{fig:admittance}d shows the QRD admittance and the total spatiotemoral admittance for discrete times $t = 0$ to $t = T_\mrm{m}/2$ where $T_\mrm{m}$ is the period of the modulation. The incident wave is assumed to be normal to the diffuser surface and $f_0 =5$~kHz. Considering the normal incidence and the geometrical symmetry of the sound diffuser about $x=0$, the nonreciprocity and the effects of the spatiotemporal modulation can be observed by assessing the change in the angular dependence of the scattered sound when modulating in the $+x$ or $-x$ directions, such that $+\beta_{n}$ for ``forward modulation'' and $-\beta_{n}$ for ``backward direction.''

We now consider a single frequency, single wavenumber spatiotemporal surface admittance given by
\begin{align}
    Y(x,t)=Y_{0}(x)+jY_\mrm{m}\cos(\omega_\mrm{m}t-k_\mrm{m}x),\label{admit_modfunction}
\end{align}
where $k_\mrm{m} = 2\pi / \Lambda$. The term for the modulation is assumed to be purely imaginary assuming no losses in the QRD. A purely reactive admittance modulation may be achieved by continuously varying cavity depth of an QRD element or altering the termination stiffness of the cavity. Based on Eq. ~(\ref{admit}), the modulation function in Eq.~(\ref{admit_modfunction}) can be approximated as using a Fourier series to match the expansion of the scattered and momentum equations as
\begin{equation}
    Y_\mrm{m}\cos(\omega_\mrm{m}t-k_\mrm{m}x) = \frac{Y_\mrm{m}}{2}\sum_{\substack{p=-1 \\ p\neq0}}^{1} \hat{\chi}^{p}(x)e^{jp\omega_\mrm{m}t},\label{mod_sine}
\end{equation}
where
\begin{eqnarray}
    \hat{\chi}^{-1}(x)&=&e^{j\frac{2 \pi}{\Lambda}x},\quad\quad p=-1,\nonumber\\
    \hat{\chi}^{1}(x)&=&e^{-j\frac{2 \pi}{\Lambda}x},\quad\quad p=1,\nonumber\\
    \hat{\chi}^{p}(x)&=&0,\quad \mathrm{elsewhere}.\nonumber
\end{eqnarray}
The surface admittance of the unmodulated QRD elements, $Y_{0}(x)$, can also be represented as a Fourier series\cite{berkhout1979theory,mechel1995wide} expansion on the spatial $2\pi n/\Lambda$ as shown in Fig.~\ref{fig:admittance}b using the following expression
\begin{equation}
    Y_{0}(x) = \sum_{n=-\infty}^{\infty}y_{0,n}\,e^{-j\frac{2\pi n}{\Lambda}x}.\label{Fourier_adm_QRD}
\end{equation}
Assembling Eqs.~(\ref{mod_sine}) and (\ref{Fourier_adm_QRD}), we can write the full expression for the modulated surface admittance required to establish conservation of momentum at $z=0$. We substitute the pressure from the impedance relation in Eq.~(\ref{momentum_imp}) for the normal velocity in Eq.~(\ref{momentum_imp}). The resulting momentum equation is then rewritten as
\begin{equation}
    - \frac{1}{c_{0}}\frac{\partial}{\partial t}\left(Y(x,t)p(x,z=0,t)\right)=-\frac{\partial }{\partial z} \left(p(x,z=0,t)\right),\label{momentum1}
\end{equation}
where the surface admittance and pressure are Fourier expansions. In order to calculate the scattered sound field, we must truncated the number of terms of the diffraction modes, $n$, and the modulation harmonics, $p$ to lie within $n \in [-N,\,\,N]$ and $p\in[-P,\,\,P]$. Then, using the truncated series and the expression for the scattering amplitude from the modulation provided in Eq.~(\ref{amp_harmonic}), we find a matrix equation for the $(2N+1)(2P+1)$ scattering amplitudes which is provided in Eq.~(S14) of the supplementary material\cite{supplement}. In this derivation, $N$ was chosen as twice of the number of elements in one period for the convergence of the expansion of the original normalized admittance provided in Eq.~(\ref{Fourier_adm_QRD})\cite{mechel1995wide,cox2016acoustic}. In a similar way, the number of the modulation harmonics was truncated at $P =10$ which is selected to include all scattering amplitudes above $-50$~dB relative to the maximum amplitude. The modulation frequency and amplitude were selected to be $f_\mrm{m}=530$~Hz and $Y_\mrm{m}=1.3$, respectively, to show the improvement of the diffusion performance.

\begin{figure*}
    \centering
    \includegraphics[width=1.0\textwidth]{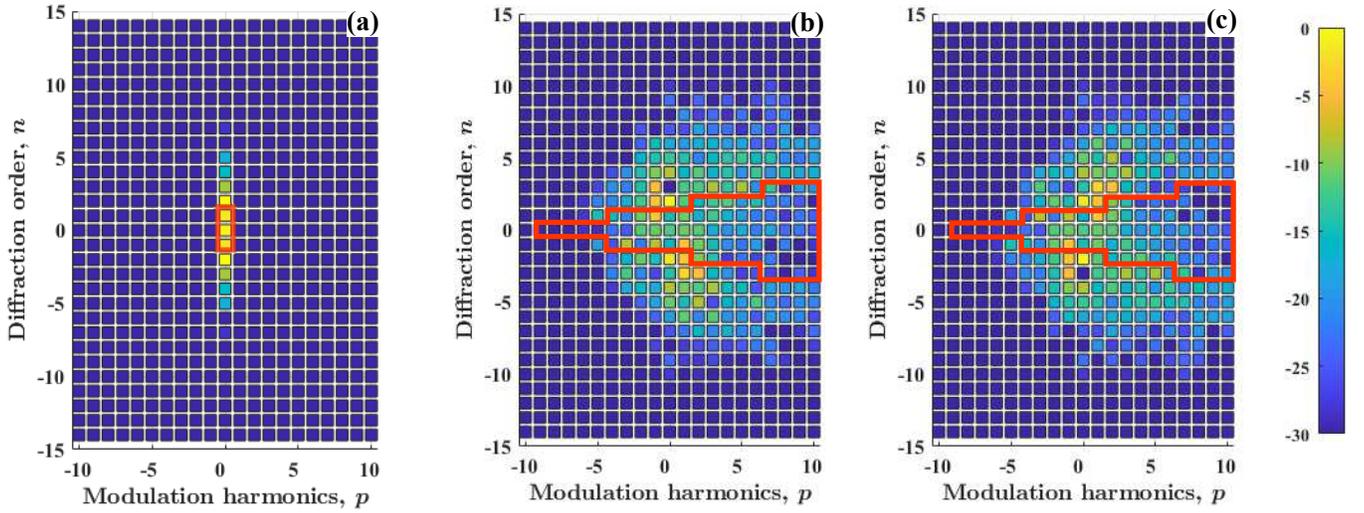}
    \caption{\label{fig:amp} Square of the scattering amplitudes normalized by the maximum scattering amplitude for all $n$ and $p$, $|{a}_{\rm n}^{\rm p}|^2/\{\mrm{max}\left[{a}_{\rm n}\right]\}^2$, for (a) unmodulated QRD, (b) backward modulation, and (c) forward modulation. Propagating modes are enclosed by the solid red lines. Color scale in decibels.}
\end{figure*}

\begin{figure}
    \subfloat[Backward modulation\label{Backward}]{
        \includegraphics[width=0.4\textwidth]{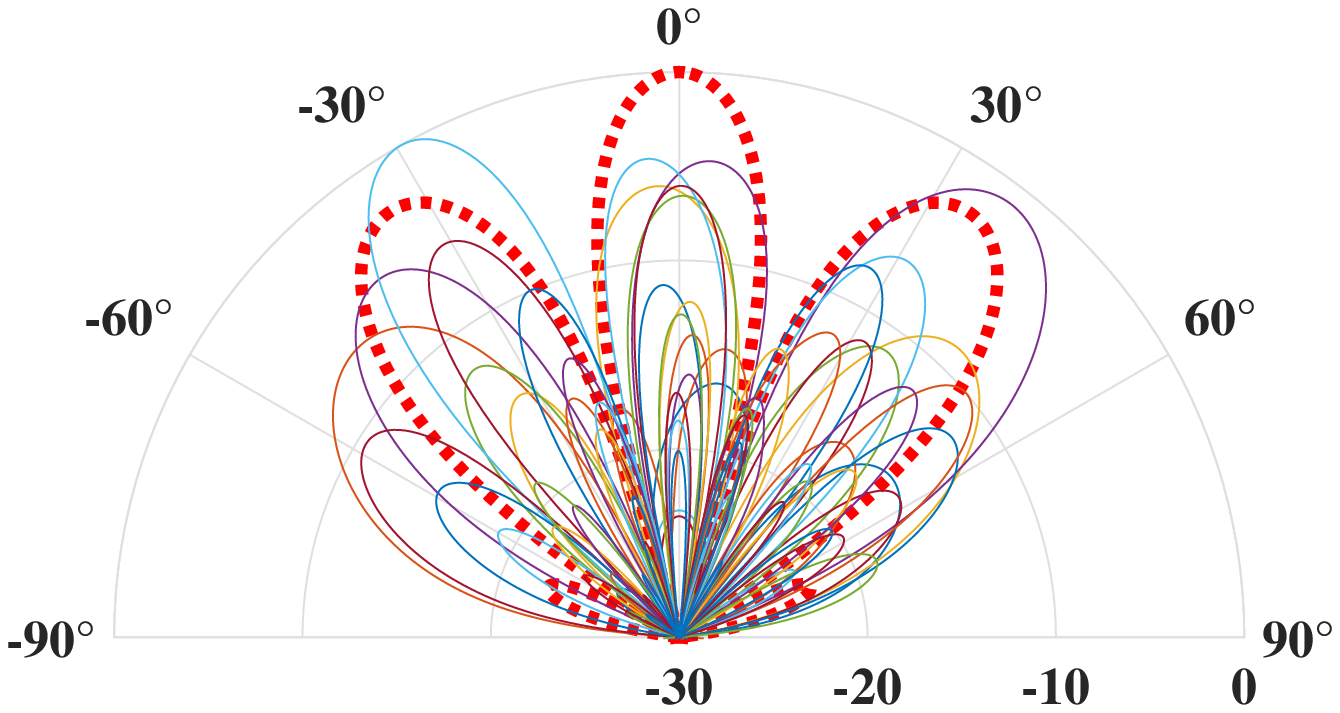}
    }\\
    \subfloat[Forward modulation\label{Forward}]{
        \includegraphics[width=0.4\textwidth]{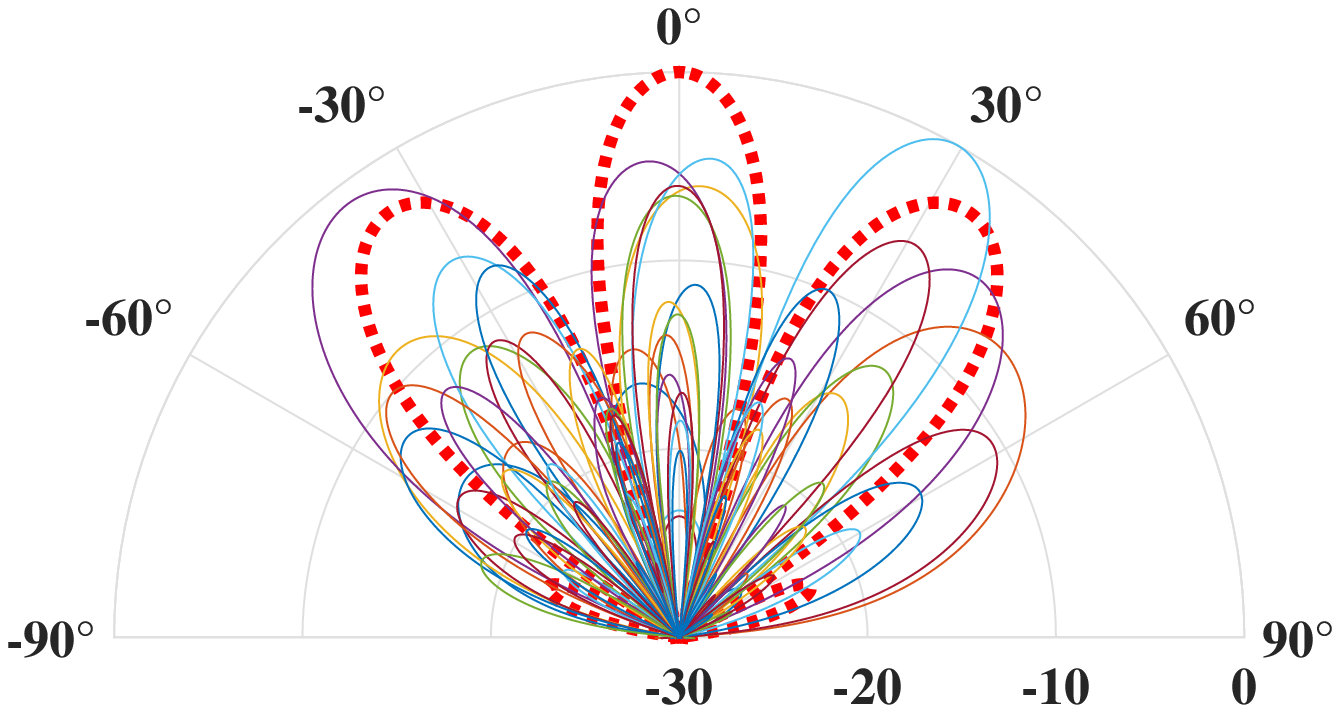}
    }
    \caption{Scattered fields for all modulation harmonics, $p\in[-P, P]$, for forward and backward modulation. The thick red dotted lines for unmodulated QRD.}
    \label{fig:all harmonic}
\end{figure}

\begin{figure*}
    \subfloat[Analytical model\label{analytical_all}]{
        \includegraphics[width=0.43\textwidth]{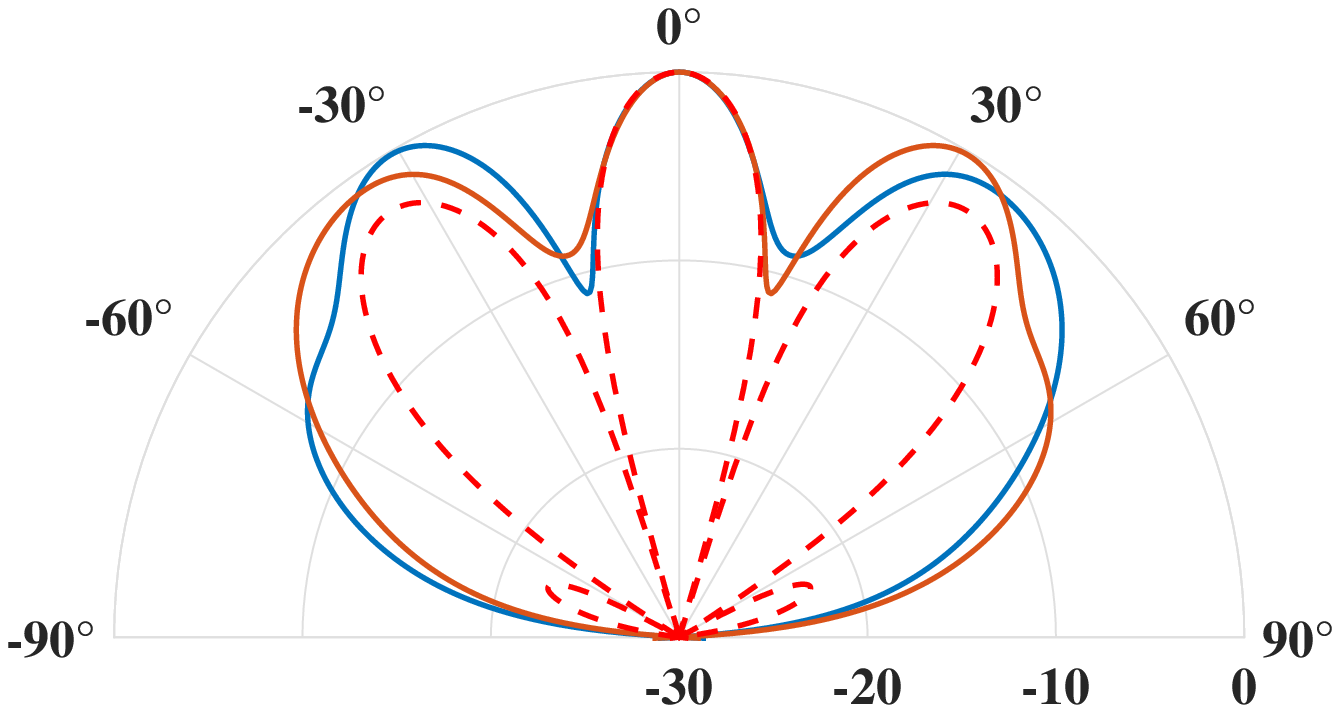}
    }\hfill
    \qquad
    \subfloat[Numerical simulation\label{Comsol_all}]{
        \includegraphics[width=0.43\textwidth]{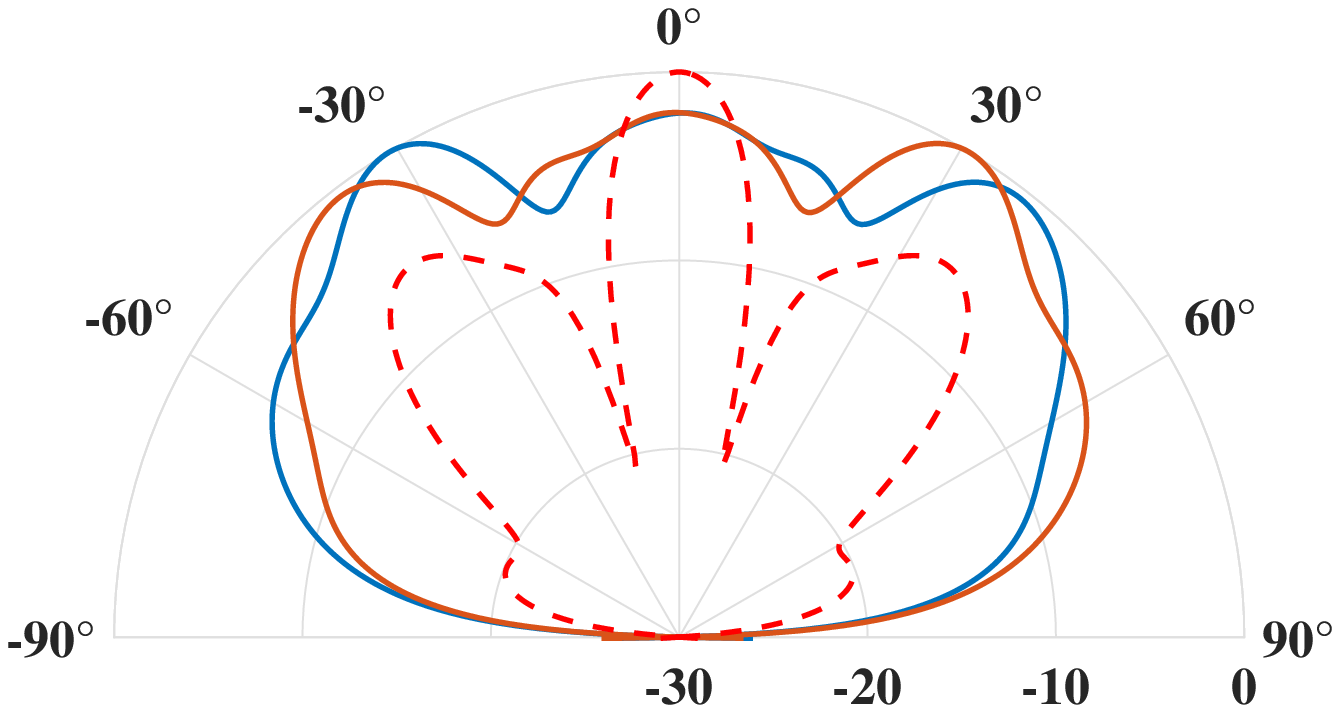}
    }
    
    \subfloat[Scattered field from numerical simulation: Modulated\label{Sim_mod}]{
        \includegraphics[width=0.43\textwidth]{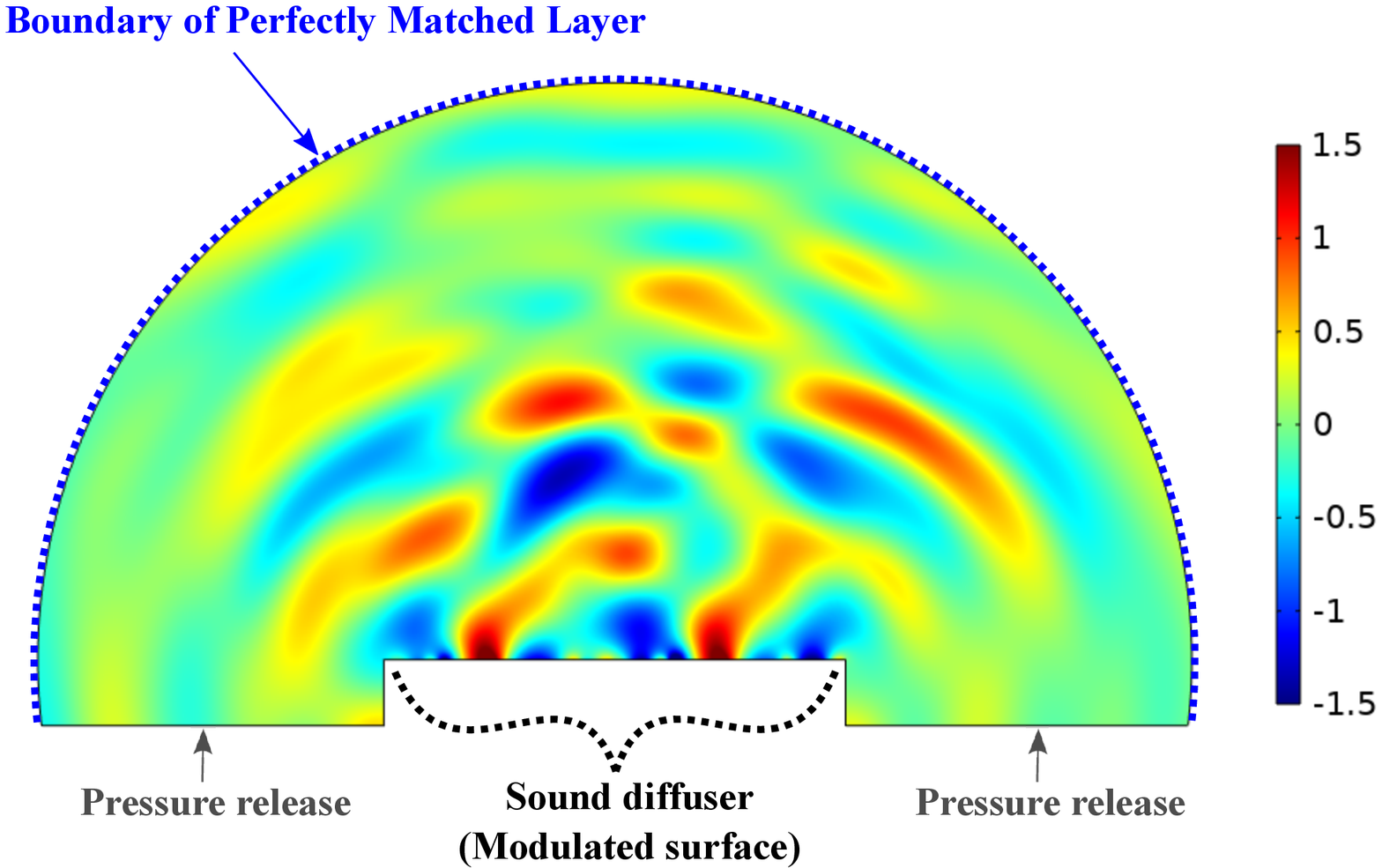}
    }\hfill
    \qquad
    \subfloat[Scattered field from numerical simulation: Unmodulated\label{Sim_nomod}]{
        \includegraphics[width=0.43\textwidth]{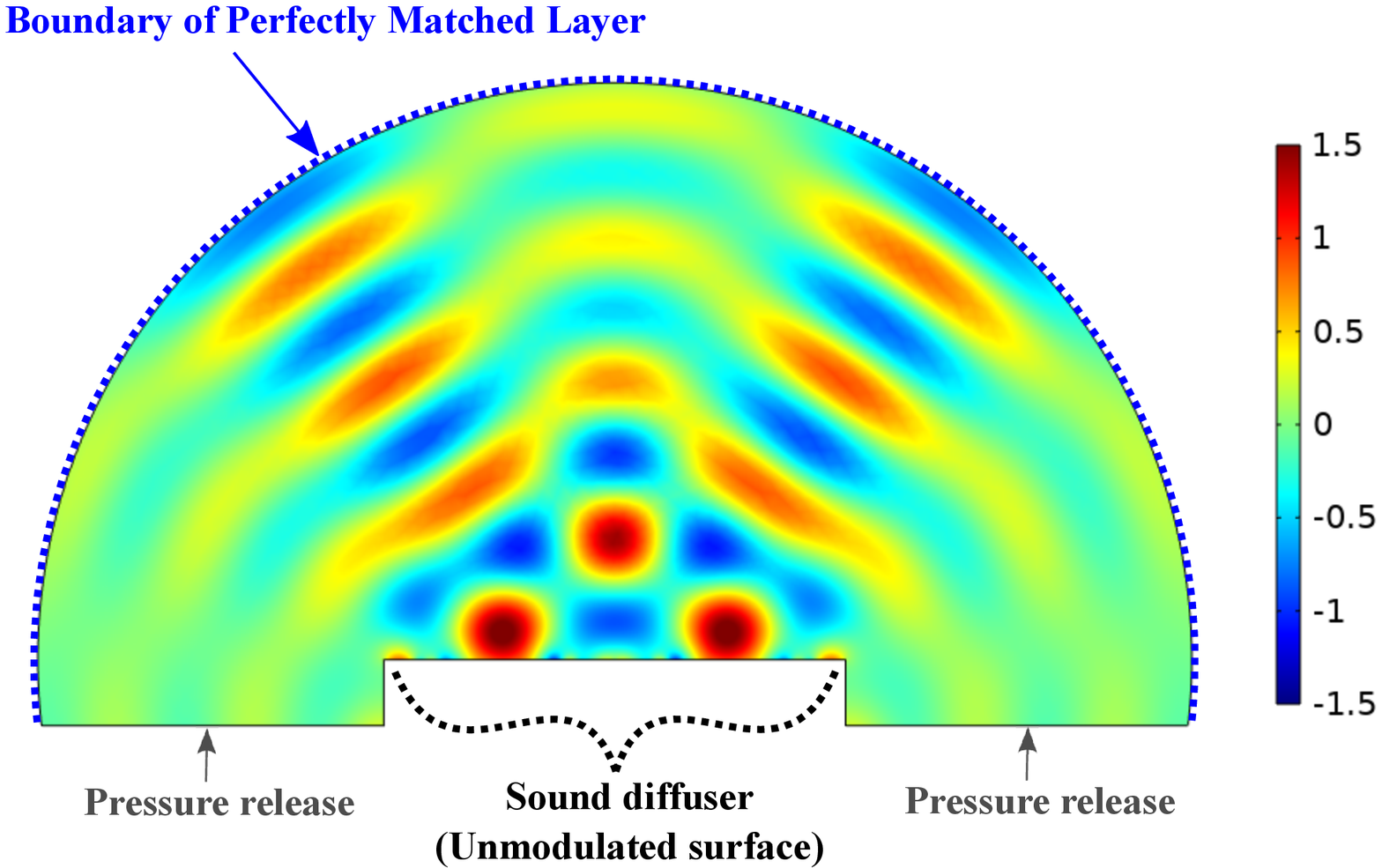}
    }
    \caption{
    Combining all responses of the modulation harmonics in (a) and of frequency bands in (b). Each curve is normalized by its maximum value. Non-modulation (red dotted line), backward (blue solid line) and forward modulation (red solid line). Transient numerical simulation for (c) the modulated surface and (d) the unmodulated surface.
    }
    \label{fig:comp_Comsol}
\end{figure*}

The magnitude of the calculated scattering amplitudes using the input parameters described here are plotted according to the indices of the diffraction modes and the modulation harmonics for two opposite spatial modulation directions in Fig.~\ref{fig:amp}.
For the unmodulated QRD, the frequency of the scattered sound is the same as the incident frequency and there are no harmonics and the only three diffracted orders propagate to the far-field, which can be observed in the three lobes in the scattering pattern for the unmodulated diffuser in Fig.~\ref{fig:all harmonic}. When spatiotemporal modulation is present, we observe non-zero scattering amplitudes at harmonics of the modulation frequency and multiple diffraction orders for each frequency. The wavenumber in the $z$-direction, $\gamma_{n}^{p}$ in Eq~(\ref{gamma}) is dependent on both the diffraction orders and modulation harmonics, so that each pair of modulation harmonics and diffracted modes represents a different direction for each component of the scattered acoustic field. As a result, each scattering amplitude contributes to the total scattered sound field, thereby distributing scattered sound power more efficiently in space when compared to the reference QRD diffuser. We can determine the scattered pressure at the surface of the diffuser using Eq.~(\ref{decomp}) and Eq.~(\ref{amp_harmonic}) to yield
\begin{equation}
    p_\mrm{s}(x,z=0,t) = \sum_{n=-N}^{N}\left(\sum_{p=-P}^{P}\hat{a}_{n}^{p}e^{j\omega_{p}t}\right)e^{-j\beta_{n}x}= \sum_{p=-P}^{P} \hat{a}^{p}(x)e^{j\omega_{p}t},\label{sc_pressure_final}\\
\end{equation}
where we have defined the spatial distribution of the scattered pressure amplitudes of the $p^{\mrm{th}}$ modulation harmonics on the diffuser surface as
\begin{equation}
    \hat{a}^{p}(x)=\sum_{n=-N}^{N}\hat{a}_{n}^{p}e^{-j\beta_{n}x}.
\end{equation}
Using the Fraunhofer far-field approximation, by taking the spatial Fourier transform of $\hat{a}^{p}(x)$, the far-field directivity for each modulation harmonic for $p \in [-P, P]$ are obtained\cite{supplement} and plotted in Fig.~\ref{fig:all harmonic}. It is observed that the locations of the grating lobes and nulls shift relative to the grating lobe directions of the unmodulated case and that the propagation directions of the diffraction modes are significantly diversified when compared to the unmodulated QRD. The generation of additional diffraction orders associated with the modulation harmonics is seen to significantly reduce minima scattered field as a function of the angle $\theta$. We note that each of the scattering patterns in Fig.~\ref{fig:all harmonic} represent the angular dependence of the propagating scattering amplitudes, each with their own shifted frequency $\omega_p$, due to a normally-incident single-frequency plane wave. To compare the response from the modulated sound diffuser with that of the unmodulated case, all modulation harmonics in Fig.~\ref{fig:all harmonic} were combined to give a single scattering amplitude at each angle since all propagating modulation harmonics contribute the scattered sound energy. Figure~\ref{analytical_all} compares the polar responses of the modulated and unmodulated QRD, showing that the spatiotemporal modulation significantly improves diffusion performance. The diffusion coefficient is used as an objective measure for assessing the diffusion performance. It is defined as in Eq.~(\ref{coeff}), which is an auto correlation for spatial similarity\cite{cox2016acoustic}. Therefore, the higher values means the better diffusion performance.
\begin{equation}
    d_{\psi}=\frac{\left(\sum^{n}_{i=1}10^{L_{i}/10}\right)^{2}-\sum^{n}_{i=1}\left(10^{L_{i}/10}\right)^{2}}{(n-1)\sum^{n}_{i=1}\left(10^{L_{i}/10}\right)^{2}}
    \label{coeff}
\end{equation}
From the polar responses in Fig.~\ref{analytical_all}, the diffusion coefficient of the forward and backward modulation for an incident frequency of 5kHz are calculated as 0.75, while the original QRD's coefficient is 0.53. These numbers clearly show the improvement of the diffusion performance by spatiotemporal modulation.

For the verification of the developed semi-analytical model, the numerical finite element model was established in Comsol including the modulated surface admittance of Eq.~(\ref{admit_modfunction}). The geometry of the numerical model is the same as the image in Fig.~\ref{fig:admittance}a and also has two periods of elements. By using the transient acoustic module, the far-field polar responses for all frequency bands including the incident frequency were obtained for the modulation and non-modulation cases\cite{supplement}. By combining all those responses, the far-field polar responses from the numerical simulation in Fig.~\ref{Comsol_all} show that the diffusion from the modulated QRD has been improved compared to the original one. The diffusion coefficient for the modulation case is 0.83, and 0.57 for the original QRD in the numerical simulation. The improved diffusion performance due to the sptiotemporal modulation can also be observed in the FEM simulation. From the transient numerical simulation in a realistic condition\cite{supplement}, the scattered sound fields at $t=T_{\rm m}/2$ for both cases were plotted in Fig.~\ref{Sim_mod} and Fig.~\ref{Sim_nomod}, which also shows more dispersed scattered field from the modulated sound diffuser compared to the three distinct propagation directions of diffraction modes from the original QRD. The images of time series were also recorded as an animation\cite{supplement}, in which the true temporal scattered field can be observed.\\

The semi-analytical model for the spatiotemporal modulation has been developed by representing the surface admittance and the scattered amplitudes with the expansion of the modulation harmonics. For the demonstration, the scattering amplitudes for multiple diffraction orders and modulation harmonics were calculated for a representative QRD geometry along with a specified modulation frequency and amplitude. The presence of modulation harmonics lead to a shift in the frequency of the scattered acoustic field which affects the propagation direction for each of the diffraction orders which significantly improves diffusion performance. The resulting diversity in scattering directions in the presence of spatiotemporal modulation leads to a significantly more uniform far-field scattering pattern when compared to the unmodulated case. This approach can therefore mitigate the minima and maxima in the scattering directivity present in conventional diffuser designs. The validity of the semi-analytical model was demonstrated by comparison with the far-field scattering response obtained using finite element methods, which showed very similar directivity and improved diffusion behavior when spatiotemporally modulating the input admittance of a conventional diffuser design.

\begin{acknowledgements}
    JK acknowledges support from Samsung Electronics. MRH acknowledges partial support from NSF EFRI Award No. 1641078.
\end{acknowledgements}

\bibliography{Modulation}


\end{document}